\documentclass[prl,aps,superscriptaddress,floatfix,tightenlines,showpacs,twocolumn,longbibliography]{revtex4-1}

\usepackage{graphicx}
\usepackage{dcolumn}   
\usepackage{bm}        
\usepackage{amssymb}   
\usepackage{amsmath}
\usepackage{url}
\usepackage{layout}
\usepackage{epsfig}
\usepackage{graphicx}
\usepackage{epstopdf}
\usepackage{booktabs}
\usepackage{float}
\usepackage[hyperindex,breaklinks]{hyperref}
\usepackage{color}

\begin{document}
\title{Effect of Disorder in a Three-Dimensional Layered Chern Insulator}

\author{Shang Liu}
\affiliation{School of Physics, Peking University, Beijing 100871, China}
\author{Tomi Ohtsuki}
\affiliation{Department of Physics, Sophia University, Chiyoda-ku, Tokyo, 102-8554, Japan}
\author{Ryuichi Shindou}
\email{rshindou@pku.edu.cn}
\affiliation{International Center for Quantum Materials, Beijing, 100871, China}
\affiliation{Collaborative Innovation Center of Quantum Matter, Beijing, 100871, China}
\begin{abstract}
We studied effects of disorder in a three dimensional layered Chern insulator. By
calculating the localization length and density of states numerically, we found two 
distict types of metallic phases between Anderson insulator and Chern insulator; 
one is diffusive metallic (DM) phase and the other is renormalized Weyl semimetal 
(WSM) phase. We show that longitudinal conductivity at the zero energy state 
remains finite in the renormalizd WSM phase as well as in the DM phase, 
while goes to zero at a semimetal-metal quantum phase transition point between 
these two. Based on the Einstein relation combined with the self-consistent Born 
analysis, we give a conductivity scaling near the quantum transition point.  
\end{abstract}

\maketitle

\emph{Introduction.} ---
During the last decade, huge research effort has been made on
topological phases in condensed matter physics \cite{1,2}.
Among many others, the critical nature of the quantum phase transition
between the  topological phase and nontopological phase is one of the most fundamental
research issues in the studies of topological phases.  A
classic example is the localization problem in the two-dimensional (2D)
integer quantum Hall (QH) phase~\cite{3} (or so-called 
`Chern insulator' ~\cite{4,4a,5}), 
where a bulk delocalized state is universally observed between topologically distinct integer 
QH phases ~\cite{6,7,8}.
The existence of the bulk delocalized state comes from 
the stability of topological edge modes
in respective integer QH phases ~\cite{9}, while its `zero-measure' property is
compatible with the scaling theory of the Anderson localization ~\cite{10}.
In fact, weak interlayer couplings among 2D QH layers make
the QH quantum critical point into a finite width of 3D metallic phase ~\cite{11,12}.

In theoretical studies, preceding efforts demonstrated that multiple-layered
Chern insulators (CI) result in a novel semimetal phase, dubbed as 
`Weyl semimetal' \cite{13,14,15,16}, 
where a pair of band touching points appear in the 3D
momentum space. The pair can be regarded as topological defects having
opposite magnetic charges, so to speak, `magnetic monopole' and `antimonopole'.
On increasing interlayer couplings, the monopole in relative to the antimonopole
winds up the 3D Brillouin zone at least once, only to annihilate
with the antimonopole again. This pair creation and annihilation process 
changes a global topological feature encoded in electronic Bloch wavefunctions, which
`allows' the system to enter into another kind of topological phase or non-topological 
gapped phase. When comparing these Weyl semimetal (WSM) physics with
localization problem in integer QH physics, one may naturally ask
`how the 2D QH quantum critical point is connected with 3D WSM phase ?'
or `Can we obtain a new insight to the localization physics in
2D QH system from the viewpoint of the 3D WSM physics ?'

\begin{figure}
\centering
\includegraphics[width=1.0\linewidth]{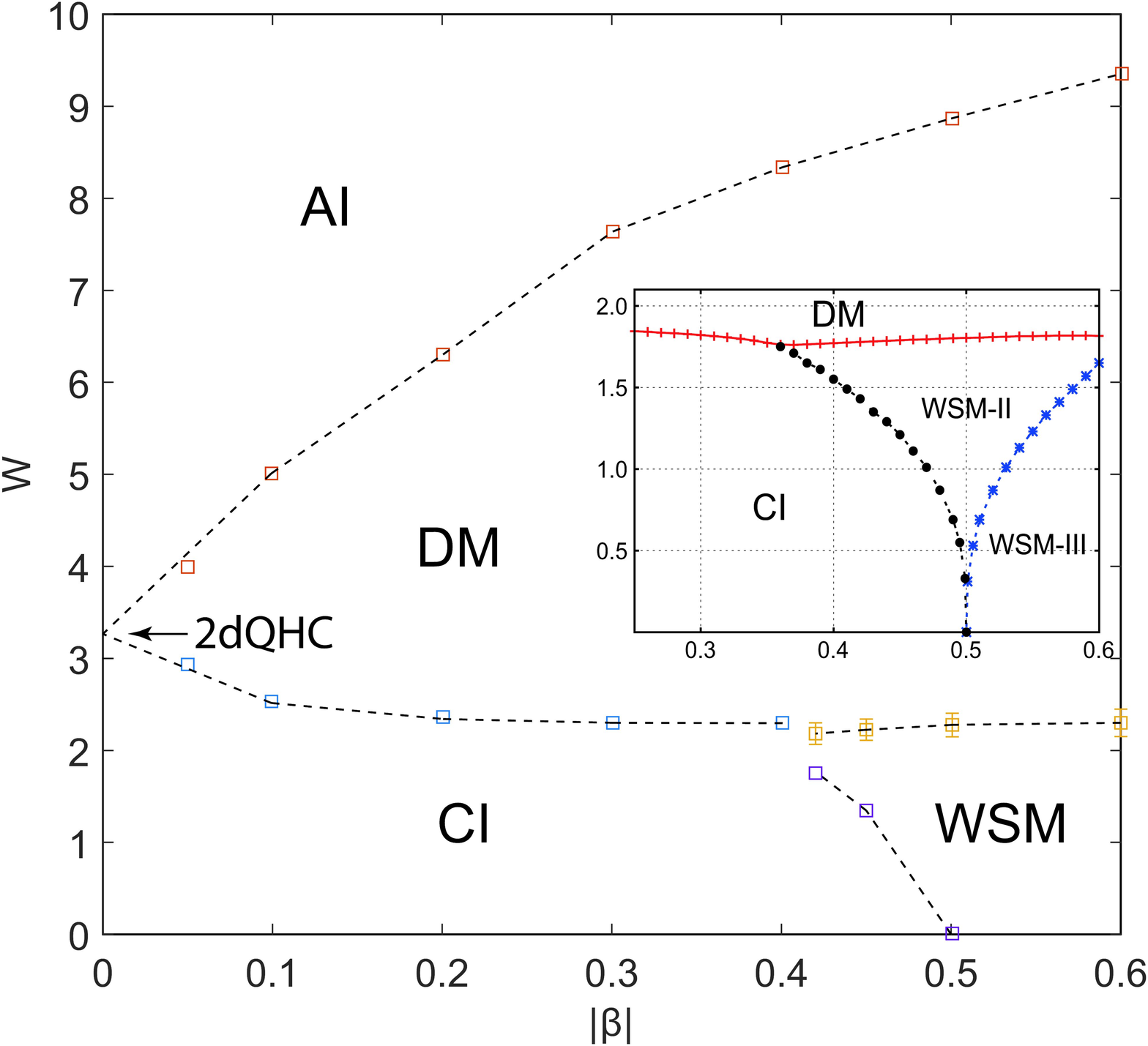}
\caption{(color online) Phase diagram of layered Chern insulator
subtended by disorder strength $W$ and the interlayer
coupling strength $|\beta|$. WSM, DM, CI, AI and 2dQHC stand for
renomarlized Weyl semimetal, diffusive metal, Chern insulator, ordinary
Anderson insulator and 2D QH quantum critical point, respectively. 
(inset) Phase diagram obtained from the self-consistent Born analysis. 
The region-II WSM phase (WSM-II) 
has two pairs of the Weyl points, while the region-III 
WSM phase (WSM-III) has three pairs of the Weyl points~\cite{19}.}
\label{fig1}
\end{figure}

In this Letter, we will answer these questions, by
studying numerically disorder effects in a layered Chern insulator.
A computation of the localization length and two-terminal
conductance by the transfer matrix (TM) method \cite{17} and density of states by the kernel polynomial expansion (KPE) method \cite{18} reveals two distinct kinds of metallic phases, which intervene
between conventional Anderson insulator (AI)
and layered Chern insulator (CI) in a phase diagram subtended by the
layer coupling and the disorder strength; one is (i) diffusive metallic (DM) phase with its zero-energy
state, i.e. an electronic state at the band touching point ($E=0$), having a finite life time and the other is
(ii) renormalized WSM phase whose zero-energy state has an infinite life time
while the velocity and location in the ${\bm k}$ space of
gapless Weyl fermions being strongly renormalized by disorder. A scaling analysis of
the density of states clarifies a critical nature of the phase transition line
between these two metallic phases. A finite-size-scaling
of the conductance also shows that finite conductivities in DM phase
and in WSM phase go to zero toward the quantum critical line.

\emph{Model.}-- We study a spinless
two-orbital tight-binding model on a cubic lattice,
which comprises of $s$-orbital and $p\equiv p_x+ip_y$ orbital;~\cite{4a}
\begin{align}
{\cal H} = & \sum_{{\bm x}} \Big(
[\epsilon_s + v_s({\bm x})] c^{\dagger}_{{\bm x},s} c_{{\bm x},s}
+ [\epsilon_p + v_p({\bm x})] c^{\dagger}_{{\bm x},p} c_{{\bm x},p} \Big)    \nonumber \\
 &+ \sum_{{\bm x}} \Big( - \sum_{\mu=x,y} \big(
t_s c^{\dagger}_{{\bm x} + {\bm e}_{\mu},s} c_{{\bm x},s}
- t_p c^{\dagger}_{{\bm x} + {\bm e}_{\mu},p} c_{{\bm x},p}\big)  \nonumber \\
& +  t_{sp} \sum_{{\bm x}}
(c^{\dagger}_{{\bm x}+{\bm e}_x,p}
- c^{\dagger}_{{\bm x} - {\bm e}_x,p})  \!\ c_{{\bm x},s} \nonumber \\
& -  it_{sp} \sum_{{\bm x}}
(c^{\dagger}_{{\bm x}+{\bm e}_y,p}
- c^{\dagger}_{{\bm x} - {\bm e}_y,p})  \!\ c_{{\bm x},s} \nonumber \\
& -  \big( t^{\prime}_s c^{\dagger}_{{\bm x} + {\bm e}_{z},s} c_{{\bm x},s}
+ t^{\prime}_p c^{\dagger}_{{\bm x} + {\bm e}_{z},p} c_{{\bm x},p}\big) + {\rm H.c.} \Big), \label{tb1}
\end{align}
where $\epsilon_s$, $\epsilon_p$ and $v_s({\bm x})$, $v_p({\bm x})$ 
denote atomic energies for the $s$, $p$ orbital and
disorder potential for the $s$, $p$ orbital, respectively.
Both $v_s({\bm x})$ and $v_p({\bm x})$ 
are uniformly distributed within $[-W/2,W/2]$ with
identical probability distribution. $t_s$, $t_p$ and $t_{sp}$ are
intralayer transfer integrals between neighboring $s$ orbitals, $p$ orbitals and that between
$s$ and $p$ orbital, respectively, while $t^{\prime}_{s}$ and $ t^{\prime}_{p}$ are 
interlayer transfer integrals. Without interlayer coupling, the model
reduces to a 2D CI, given that the so-called band inversion
condition is satisfied; $0<|\epsilon_s-\epsilon_p|<4(t_{s} + t_{p})$.
In the rest of this Letter, we take $\epsilon_s-\epsilon_p=-2(t_s+t_p)$, while
$t^{\prime}_s=-t^{\prime}_p>0$ and $t_s=t_p>0$, $t_{sp}=4t_s /3$ with $4t_s$ 
being the energy unit.~\cite{19} This parameter set realizes
a CI with a large band gap in the 2D limit
($\beta \equiv \frac{t^{\prime}_p-t^{\prime}_s}{2(t_s+t_p)} =0$) for
the half filling case. When $0\le |\beta|<1/2$, the system is fully gapped, which
belongs to the CI phase in the 2D limit. When $|\beta|>1/2$,
the system enters into 3D WSM phase, where three pairs of the monopoles and antimonopoles
appear at ${\bm k}=(0,\pi,\pi \pm k_0)$, $(\pi,0,\pi \pm k_0)$ and $(\pi,\pi,\mp k_1)$
respectively. The Weyl points at $(0,\pi,\pi-k_0)$, $(\pi,0,\pi-k_0)$ and
$(\pi,\pi,-k_1)$ are the magnetic monopoles with positive magnetic charge,
while the others are the antimonopoles with negative charge.
At $|\beta|=\frac{1}{2}$, three pairs of monopoles and antimonopoles annihilate with each other
simultaneously, i.e. $k_0=k_1=0$.

 \begin{figure}
\centering
\includegraphics[width=1.0\linewidth]{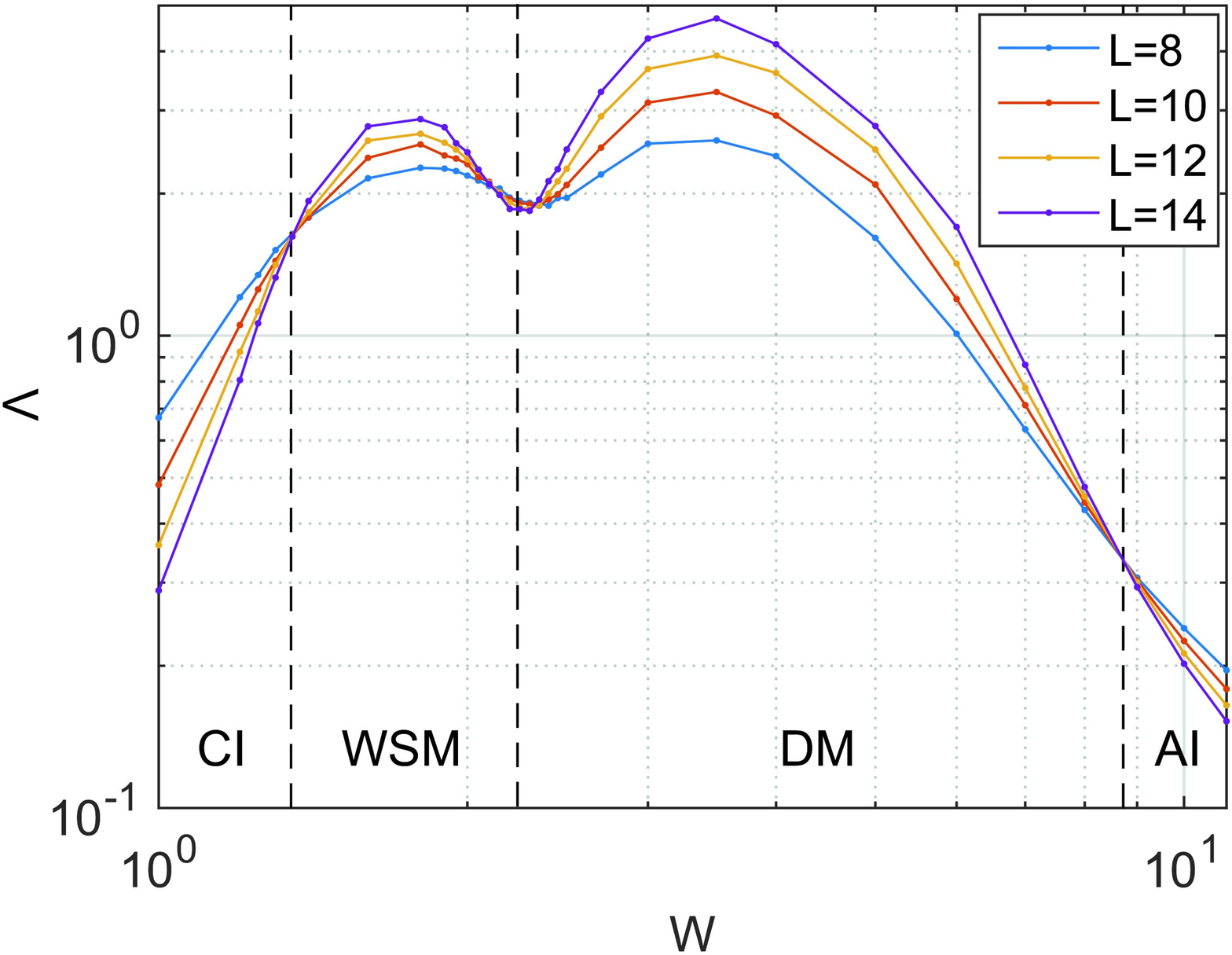}
\caption{(color online) Localization length normalized by the system size, $\Lambda$,  
as a function of disorder strength
for $|\beta|=0.45$ and various system size $L$. For a guide to the eyes, we put three vertical
dotted lines for the scale-invariant critical points where the localization length barely changes with $L$.
We note that the system is highly anisotropic, and the transfer matrix calculation along $x$-direction 
(in-plane direction) seems to suffer large corrections to scaling.}
\label{fig2}
\end{figure}

\emph{Localization Length and Phase Diagram.}---
By the TM method, the localizaiton length along the $z$-direction (the stacking direction) is
calculated for various system sizes ($N \equiv L^2\times 10^{5} \sim 10^6$ with
$L=8,10,12,14$,  $L$ being the linear dimension of the cross section)
as a function of $W$ and $|\beta|$. In the presence of finite
interlayer coupling ($|\beta| \ne 0$), the QH critical point in the 2D
limit becomes a finite window of a metallic region between AI and
CI. With larger coupling, the finite region of
the metallic phase become two distinct metallic phases, which are separated
by a critical line.
One is in stronger disorder side and is connected with
the 2D QH critical point, while the other is in weaker
disorder side and belongs to the same phase as the 3D WSM phase
in the clean limit (Fig.~\ref{fig1}).
These two metallic phases are always separated by a transition line where the
localization length $\xi$ normalized by the linear dimension of the
system size $L$ barely changes as a function 
of $L$ (Fig.~\ref{fig2}) \cite{20,21,22}.The scale-invariant
behaviour of $\Lambda \equiv \xi/L$ suggests existence of
quantum critical line between these two metallic phases.

\begin{figure}
\centering
\includegraphics[width=1.0\linewidth]{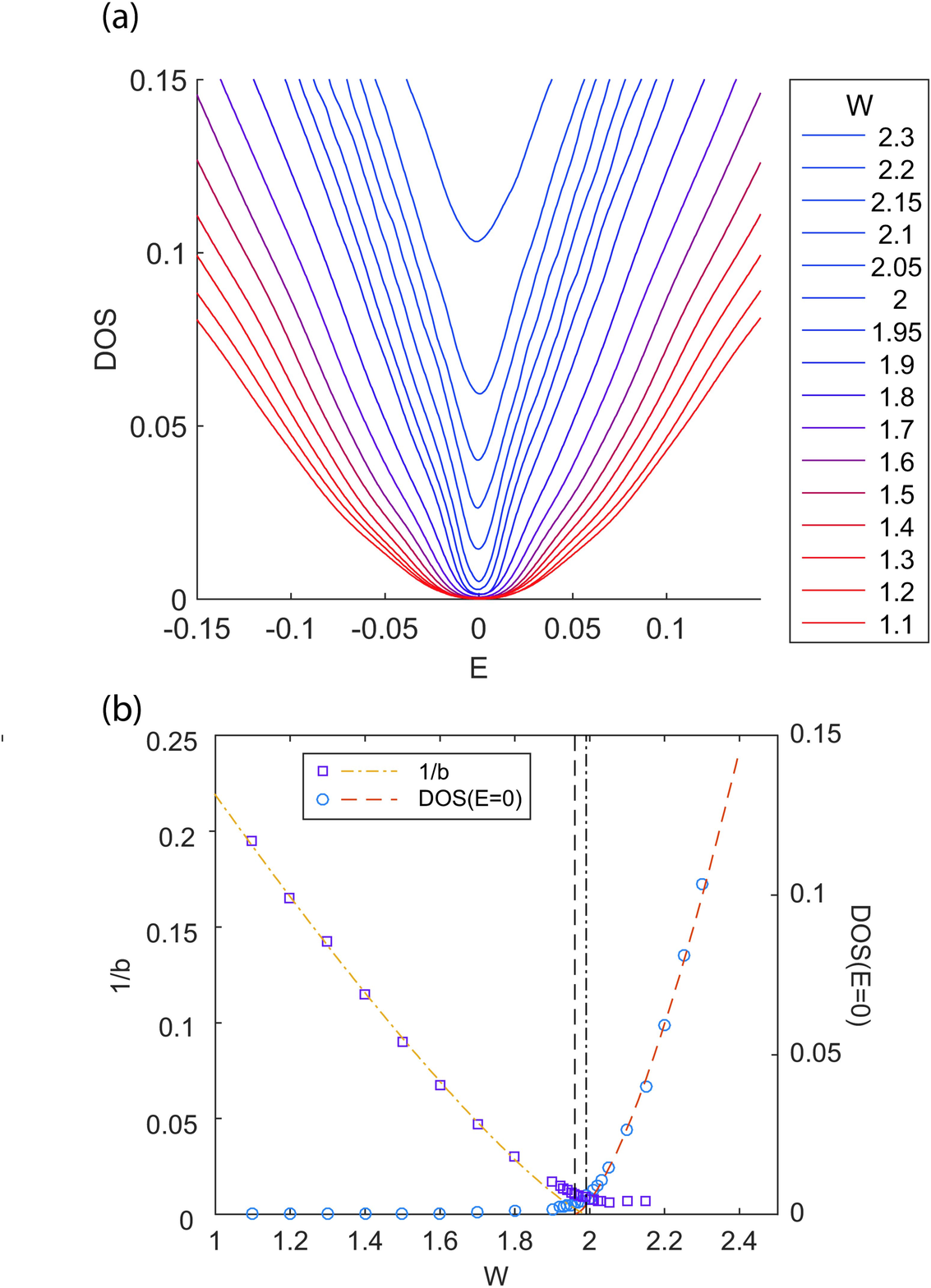}
\caption{(color online) (a) Density of states (DOS) at $|\beta|=0.6$ near the zero energy state
for those $W$ near the critical point intervening between
DM and WSM phase. $W$ increases from bottom to top.
(b) Zero-energy density of states $\rho(E=0)$ and cube of averaged velocity $\overline{v}$ 
as a function of $W$. The latter is evaluated from the fitting of $\rho(E)$ to $a+bE^2$ 
with $\overline{v}^3\propto 1/b$ and free parameter $a$ (see also text). The vertical
dot-dashed line and dashed line denote the critical disorder
strength at which $\overline{v}$ and $\rho(E=0)$ vanish, respectively (see
Ref.~\onlinecite{19} for their determinations).}
\label{fig3}
\end{figure}

\emph{Density of States.}---To characterize this critical line,
we computed the density of states (DOS) in terms of the KPE method for several
different cubic system sizes ($N=L^3$ with $L=40,~48,~60~\text{and}~80$), to extrapolate
the thermodynamic behaviour of the DOS based on
$\rho_{L}(E,W)=\rho(E,W)+b(E,W)/L^2$ \cite{19}. Here $\rho_{L}(E,W)$ denotes
the DOS as a function of the electron energy $E$ and $W$ for the linear dimension
$L$. The intercept $\rho(E,W)$ as a function of $1/L^2$ can be regarded as
the density of states in the thermodynamic limit.
Fig.\ref{fig3} shows the DOS near the zero energy for those
disorder strengths around the phase transition between the two metallic
phases. The DOS for the weaker disorder side vanishes at the zero energy,
where $\rho(E,W)$ becomes a parabolic function in $E$. The DOS in the stronger disorder
side acquires a finite value at $E=0$ \cite{23,24,25,26}. 
This feature is consistent with self-consistent Born analyses 
(SCBA) \cite{19,24,27,27a} and renormalization 
group (RG) analysis \cite{23,28}, suggesting that the former metallic phase 
belongs to the same phase as the WSM phase in the clean limit; renormalized 
WSM phase whose zero-energy state (an electronic state at the band 
touching point; $E=0$) has an infinite life time. Meanwhile the latter phase is 
characterized by the zero-energy state having a 
finite life time; diffusive metallic (DM) phase.

In the renoarmlized WSM phase, the density of states near $E=0$ is determined by
an effective Hamiltonian linearized near the gapless points, 
$H_{\mathbf{k}}=\sum_iv_ip_i\sigma_i$, where $\mathbf{v}$ is a renormalized velocity and
$\mathbf{p} \equiv \mathbf{k}-\mathbf{k}_0$ the momentum distance from
the respective Weyl point $\mathbf{k}_0$. Each Weyl point contributes
to DOS near $E=0$ as
$\rho(E)=\frac{1}{L^3} \sum_{\mathbf{p}} \delta\left(E-E_\mathbf{p}\right)=\frac{1}{2\pi^2 {\bar v}^3}E^2$,
with $\bar v\equiv |v_xv_yv_z|^{1/3}$ being an averaged velocity.  
On increasing the disorder strength, the renormalized velocity $\overline{v}$ 
evaluated from the coefficient of $E^2$ decreases toward a certain critical point ($W=W_c$); 
purple squares in Fig.~\ref{fig3}b. At the same point, $\rho(0)$ starts to take a 
finite non-negligible value; blue circles in Fig.~\ref{fig3}b, where the 
SCBA identifies $\rho(0)$ as inversely proportional to the life time of the 
zero-energy state~\cite{19}. We also find that 
all the data points for the density of states near $E=0$ collapse into 
a single-parameter scaling function form, $\rho(E) = \delta^{(3-z)\nu}f(|E|\delta^{-z\nu})$, 
with $\delta\equiv |W-W_c|/W_c$~\cite{25,29}. By fitting the DOS curves into 
this scaling function, we determine the dynamical exponent $z=1.53\pm 0.03$ 
and $\nu=0.84\pm 0.1$.~\cite{19}  
The scaling analysis in combination with the SCBA unambiguously conclude the 
existence of the quantum critical line between DM phase and WSM phase at 
$W=W_c$. 

\begin{figure}
\centering
\includegraphics[width=1.0\linewidth]{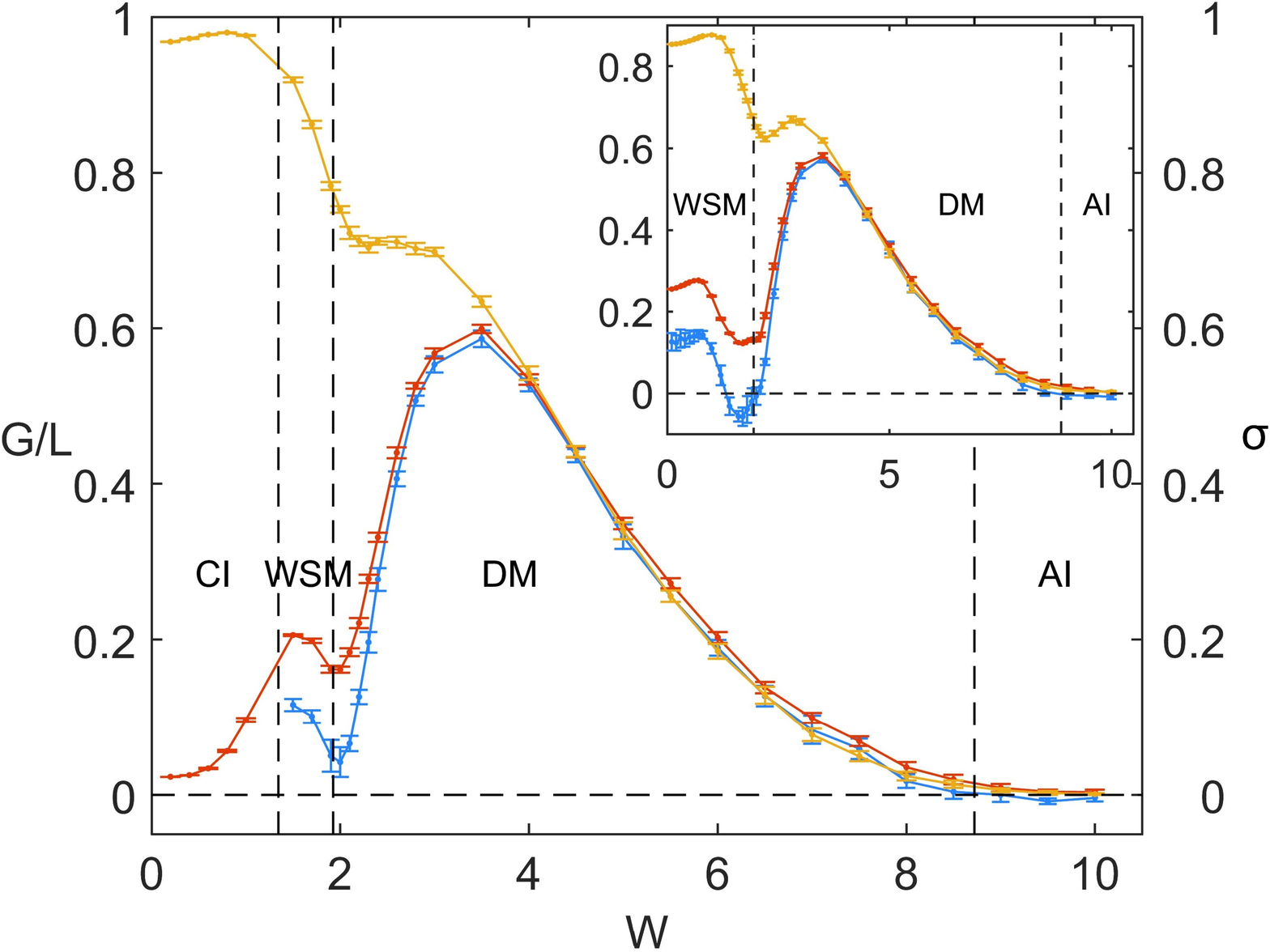}
\caption{(color online) Bulk conductivity  and conductances 
of the cubic system ($L^3$) at $|\beta|=0.45$ (inset: at $|\beta|=0.50$)  
as a function of disorder strength $W$; 
the bulk conductivity $\sigma_b$ (blue points with line), 
conductance with the periodic boundary condition $G^p$ ($L=30$; red points with line) and that 
with open boundary condition $G^o$ ($L=30$; yellow points with line). The vertical
dotted lines around $W=1.92$ ($|\beta|=0.45$) and  $W=1.96$ ($|\beta|=0.50$)  
stand for the WSM-DM transition points determined by the DOS,
while the vertical dotted lines around $W=8.7$ ($|\beta|=0.45$)  and $W=8.9$ ($|\beta|=0.50$) 
and that around $W=1.35$ ($|\beta|=0.45$) are the DM-AI transition points and CI-WSM transition point 
determined from the localization length, respectively.}
\label{fig4}
\end{figure}
\begin{figure}
\centering
\includegraphics[width=1.0\linewidth]{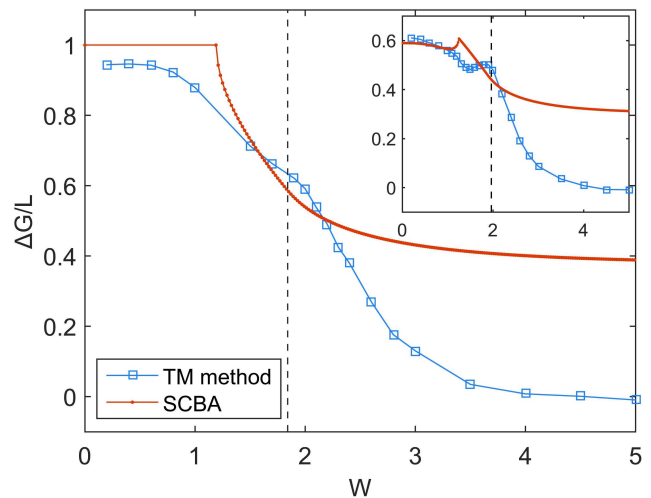}
\caption{(color online) $\Delta G \equiv G^{o}-G^{p}$ for $|\beta|=0.45$ (inset for 
$|\beta|=0.55$), where $G^o$ and $G^p$ are calculated for the cubic system 
($L^3=30^3$). The red solid line is a surface conductance 
evaluated from eq.~(\ref{surfaceG}) along with the SCBA result for the same 
parameter set.}  
\label{fig5}
\end{figure}
\emph{Two-Terminal Conductance and Conductivity} ---
To characterize the renormalized WSM and DM phases,
we calculated the two-terminal conductance $G$ in the in-plane direction ($x$-direction)
for the cubic system size with various
linear dimensions ($L^3$ with $L=6,8,\cdots,30$)
via the transfer matrix method \cite{31}. In the
other two spatial directions ($y$ and $z$-directions),
we impose either periodic boundary condition ($G^p$)
or open boundary condition ($G^{o}$).
In the Chern insulator (CI) phase, $G^p$ is nearly zero at
$L=30$, while, for the two metallic phases, an in-plane bulk conductivity $\sigma_b$ is
obtained from $G^p$ by the linear fitting ($G^p=\sigma_bL+b$). 

The calculated result shows that finite conductivity in the DM and the 
WSM phase reduces to zero toward $W=W_c$ (Fig.~\ref{fig4}). 
When combined with the SCBA result, this behaviour is consistent with the Einstein relation 
for the conductivity $\sigma(E)$; 
\begin{eqnarray}
\sigma(E) =e^2  D(E)\rho(E).  \label{einstein}
\end{eqnarray}
The diffusion constant $D(E)$ is given by the life time $\tau$ and the averaged velocity 
$\overline{v}$ as $D(E)=\frac{\overline{v}^2\tau}{3}$. SCBA relates $\rho(E)$ with the inverse 
of the life time as $\rho(E) = 24/(\pi \tau W^2)$~\cite{19}. Thus the relation 
tells that $\sigma(E=0)=8 e^2 \overline{v}^2/(\pi W^2)$ remains finite not only in the DM phase 
but also in the renormalized WSM phase. Moreover, since $\overline{v}$ vanishes at 
the critical point as above~\cite{25}, 
the Einstein relation also dictates 
that the conductivity reduces to zero at the critical point,
being consistent with our numerical observation 
(blue points near $W=2$ in Fig.~\ref{fig4}).  Note also that $\sigma(E=0)$ vanishes 
as $\delta^{(d-2)\nu}$ from the DM side while $\overline{v}$ vanishes as $\delta^{(z-1)\nu}$ 
from the WSM side~\cite{25}. With our expression for the conductivity 
$\sigma(E=0) \propto \overline{v}^2$, we see that the conductivity vanishes continuously 
from the WSM side as $\sigma(E=0) \propto \delta^{2(z-1)\nu}$. This is qualitatively consistent 
with Ref.~\onlinecite{31a}, which predicts a continuously vanishing conductivity with a 
different exponent $\sigma(E=0)\propto \delta$. If we assume that $z=d/2$, our exponent 
in the WSM side coincides with that from the DM side.

$G^{o}$ and $G^{p}$ signficantly differ from each other in the Chern insulator
phase and the renormalized WSM phase (Fig.~\ref{fig4}).
The difference can be attributed to the surface 
conductance due to the chiral surface states. In the renormalized WSM phase, 
the zero-energy state has an infinite life time $\tau$, so that the SCBA 
is fully justified; neglected Feynman diagrams in the approximation 
are smaller than those included at least by a factor $1/\tau$. The SCBA 
result shows that the Green's function after quenched 
impurity averaging has either three or two pairs of Weyl points in the momentum 
space (inset of Fig.~\ref{fig1}). In the region-II WSM phase, two pairs of Weyl points appear at 
${\bm k}=(0,\pi,\pi\pm \overline{k}_0)$ and $(\pi,0,\pi\pm \overline{k}_0)$, while  
another one pair appears at $(\pi,\pi,\mp \overline{k}_1)$ in the 
region-III WSM phase. These Weyl points in the bulk electronic state result in a surface 
chiral Fermi arc state with left-moving chirality connecting from the surface crystal
momentum $(k_x,k_z)=(0,\pi-\overline{k}_0)$ to 
$(0,-\pi+\overline{k}_0)$, and two other chiral arc states with right-moving chirality, 
one connecting from $(\pi,\pi-\overline{k}_0)$ to $(\pi,\overline{k}_1)$ and the other 
from $(\pi,-\overline{k}_1)$ to $(\pi,-\pi+\overline{k}_0)$. They in total lead to a finite 
two-terminal surface conductance which depends on the length of each Fermi arc. 
The left-moving chiral arc and right-moving arc cancel each other by the intra-surface  
backward scattering due to impurities, so that the surface conductance is expected to be 
expressed with the difference between the length of right-moving arc ($2\pi-2\overline{k}_0-2\overline{k}_1$)  
and that of the left-moving arc ($2\overline{k}_0$);     
\begin{eqnarray}
\Delta G \equiv G^{o}-G^{p} = \frac{e^2}{h} \frac{(\pi -2\overline{k}_0-\overline{k}_1) L}{\pi}.     
\label{surfaceG} 
\end{eqnarray}
In fact, $\overline{k}_0$ and $\overline{k}_1$ obtained from the SCBA  along with eq.~(\ref{surfaceG}) 
reproduces $\Delta G$ from the numerics at the quantitative 
level for $W < W_c$ (Fig.~\ref{fig5}). Importantly, paired two Weyl nodes do {\it not} 
annihilate with each other even at $W=W_c$; the associated chiral surface Fermi arc state 
{\it does} survive up to $W = W_c$. For $W > W_c$, each Weyl node acquires a finite 
line broadening (inverse of lifetime), so that the chiral Fermi arc states near the two 
ends of the arc start to be mixed with bulk states and lose their protected surface 
conduction property. Meanwhile, those arc states away from the two ends still 
contribute to robust surface conductions for $W \gtrsim W_c$.~\cite{19} 
This picture is consistent with our numerical observation of 
the finite surface conductance $\Delta G$ 
for $W \gtrsim W_c$, which is smaller than the SCBA estimate from Eq.~(\ref{surfaceG}) 
in the DM phase. 

\emph{Conclusion}---
In this Letter, we introduce a simple lattice model in the unitary class, 
which enables us to study WSM phase, disorder-induced DM phase and quantum 
phase transition between these two phases. The critical nature of the phase transition 
is confirmed by the scale-invariant behaviour of the localization length and by 
the scaling analysis of the density of states. The calculated critical exponents 
take those values close to critical exponents previously obtained in a lattice model 
of the symplectic class~\cite{25}. We show that longitudinal conductivity at the 
zero-energy state remains finite in the renormalized WSM phase as well as in the DM 
phase, while it reduces to zero toward the transition point. With the help of the SCBA, 
we show that this observation is consistent with the Einstein relation for the conductivity 
and Wegner's conductivity scaling~\cite{32}.  The two-terminal conductance 
shows a significant residual surface conductance at the transition point, which 
can be attributed to the ``renormalized'' chiral Fermi arc states.

\emph{Acknowledge}  The authors would like to thank
Dr. Koji Kobayashi for fruitful discussions.  This work was supported by
JSPS KAKENHI Grants No.15H03700 and No. 24000013 and by 
NBRP of China (Grant No. 2015CB921104). 

\emph{Note added.} -- Recently we became aware of independent
numerical works~\cite{33} that treat similar situation but focus on a different aspect.

\section{Supplemental materials}
\subsection*{Weyl semimetal phases and chiral Fermi arc in layered Chern insulator }
\begin{figure}
\centering
\includegraphics[width=1.0\linewidth]{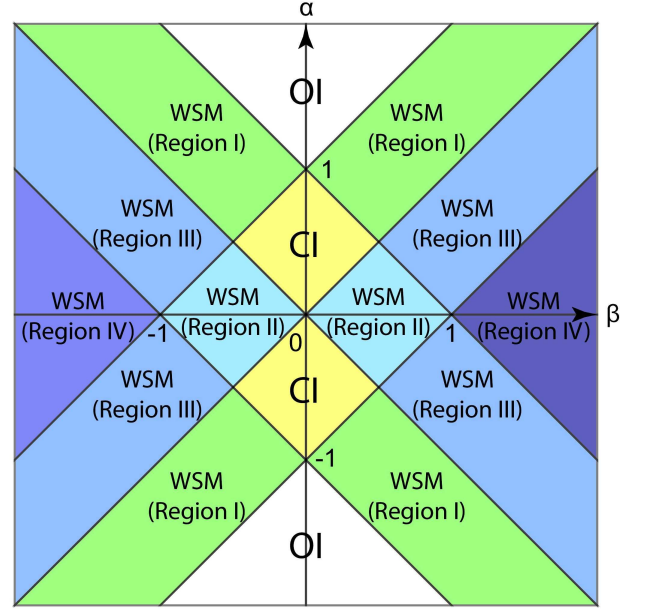}
\caption{(color online) Phase diagram of layered Chern insulator
in the clean limit. OI and CI denote ordinary band insulator and Chern insulator respectively,
while WSM denotes the Weyl semimetallic phases. The explanation for four distinct regions of
the WSM phase is given in the text of this supplemental material.}
\label{pd}
\end{figure}

\begin{figure}
\centering
\includegraphics[width=0.9\linewidth]{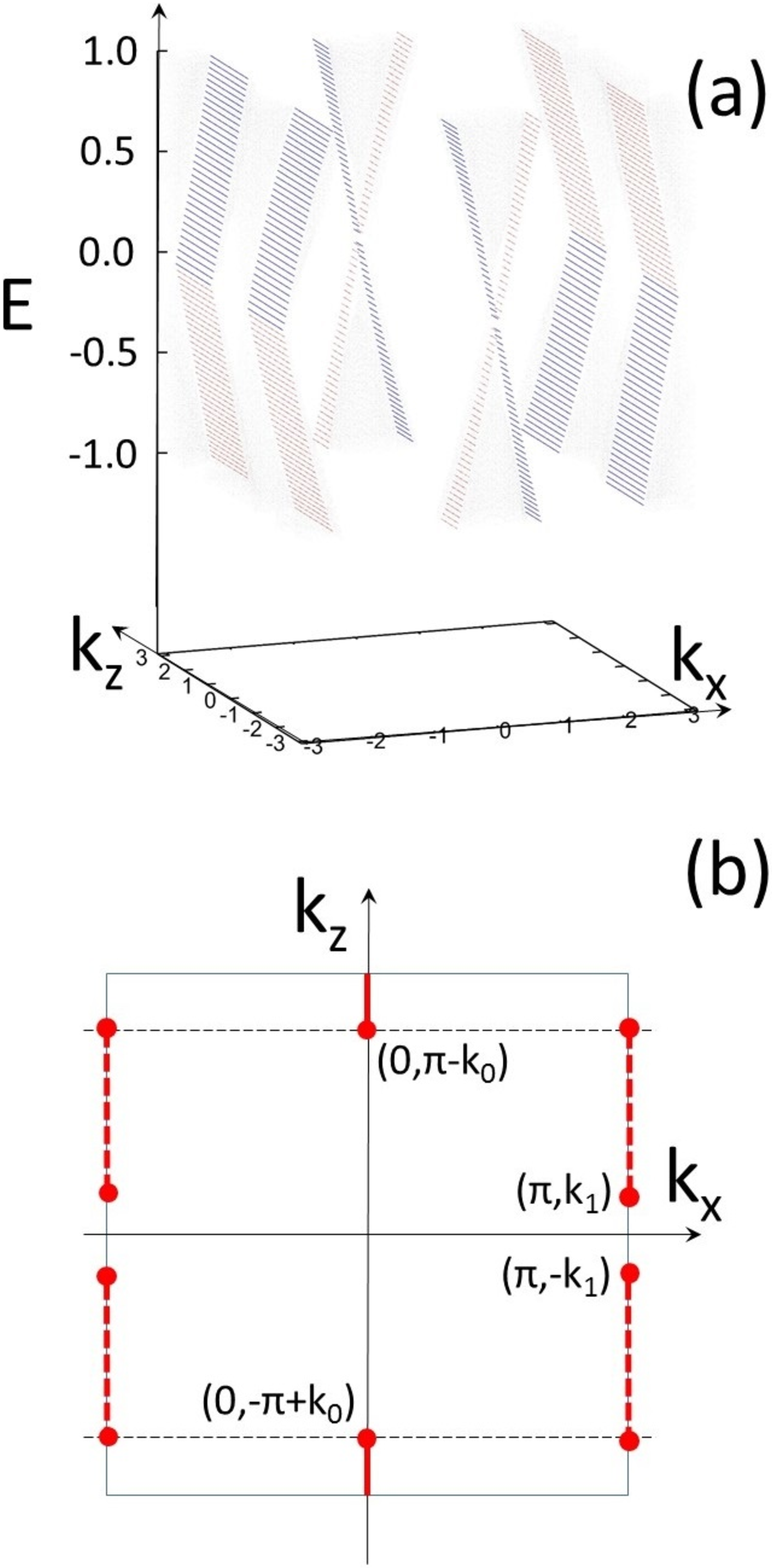}
\caption{(color online) (a) Energy dispersion of chiral Fermi surface states localized at one surface
(red) and at the other (blue). Those energy states with grey color points are bulk states.
(b) Location of chiral Fermi arcs are depicted on the surface crystal momentum space (the
$k_x$-$k_z$ plane), where a red solid line stands for a right-moving chiral Fermi surface state
($dE/dk_x>0$), and red dotted lines are for left-moving chiral Fermi surface states ($dE/dk_x<0$).}
\label{CFA}
\end{figure}

The tight-binding Hamiltonian for layered Chern insulator \cite{4,4a} 
reduces to the following 2 $\times$ 2 Hamiltonian
in the momentum space,
\begin{eqnarray}
{\bm H}({\bm k}) = a_0 \sigma_0 + {\bm a}\cdot {\bm \sigma} \label{eq1}
\end{eqnarray}
with ${\bm \sigma}=(\sigma_x,\sigma_y,\sigma_z)$ are Pauli matrices and
\begin{align}
a_0({\bm k}) &= \frac{\epsilon_s+\epsilon_p}{2} + (t_p-t_s) (\cos k_x + \cos k_y)
- (t^{\prime}_s + t^{\prime}_p) \cos k_z, \nonumber \\
a_3({\bm k}) &=  \frac{\epsilon_s-\epsilon_p}{2} - (t_p+t_s) (\cos k_x + \cos k_y)
- (t^{\prime}_s - t^{\prime}_p) \cos k_z, \nonumber \\
a_1({\bm k}) &= -2t_{sp} \sin k_y, \nonumber \\
a_2({\bm k}) &= -2t_{sp} \sin k_x, \nonumber
\end{align}
respectively. For simplicity, we take $t_p=t_s>0$ and $t^{\prime}_s=-t^{\prime}_p$ so that the
${\bm k}$-dependence of $a_0$ can be omitted and the direct band gap always becomes the
global band gap $\Delta({\bm k}) = 2\min|{\bm a}({\bm k})|$. If any, the gap closing occurs at
the high symmetric ${\bm k}$ lines, i.e. ${\bm k}=(0,0,k_z),(\pi,0,k_z),(0,\pi,k_z),(\pi,\pi,k_z)$.
At these points, the gap is given as a function of $k_z$
\begin{eqnarray}
\Delta({\bm k}) = 2(t_s+t_p) \cdot \bigg|\alpha - \frac{\cos k_x + \cos k_y}{2}
+ \beta \cos k_z\bigg|, \nonumber
\end{eqnarray}
with  $\alpha \equiv \frac{\epsilon_s-\epsilon_p}{4(t_s+t_p)}$ and
$\beta\equiv \frac{t^{\prime}_p-t^{\prime}_s}{2(t_s+t_p)} = -\frac{t^{\prime}_s}{2t_s}$.
This gives a phase diagram subtended by $\alpha$ and $\beta$ depicted in Fig.~\ref{pd}.
The phase diagram comprises of Chern insulator phases and conventional band insulator
phases, and four Weyl semimetal phases; WSM in the region I has
one pair of two Wely points having opposite magnetic charges
(either at ${\bm k}=(0,0,\pm *)$ or $(\pi,\pi,\pm *)$), WSM in the regions II
has two pairs (at ${\bm k}=(\pi,0, \pm *),(0,\pi,\pm *)$).
In the region III and VI, it has three pairs (either at ${\bm k}=(0,0,\pm *)$, $(\pi,0,\pm *)$, $(0,\pi,\pm *)$,
or at $(\pi,\pi,\pm *)$, $(\pi,0,\pm *)$, $(0,\pi,\pm *)$) and four pairs respectively.
In the main text, we took $\alpha=-\frac{1}{2}$, so that, on increasing $|\beta|$, the
system enters the region III first ($\frac{1}{2}<|\beta|<\frac{3}{2}$)
and then region IV ($\frac{3}{2}<|\beta|$).  Figure~\ref{CFA} shows surface chiral Fermi arcs
in the case of $|\beta|=0.6$, where the right-moving chiral Fermi arc connects from a surface momentum
point $(k_x,k_z)=(0,\pi-k_0)$ to $(0,-\pi+k_0)$ with $k_0 \equiv \cos^{-1}[\frac{\alpha}{\beta}]$,
while the two left-moving chiral Fermi arcs connect from $(k_x,k_z)=(\pi,k_1)$ to
$(\pi,\pi-k_0)$ and from $(k_x,k_z)=(\pi,-k_1)$ to $(\pi,-\pi+k_0)$ respectively with
$k_1 \equiv {\cos}^{-1}[-\frac{1+\alpha}{\beta}]$.

\subsection*{Self-consistent Born analysis}
A single-point Green function is averaged over the on-site impurity potentials,   
\begin{eqnarray}
[{\bm G}_{\pm}({\bm k},{\bm k}')]_{\alpha,\beta} \equiv 
\big\langle \langle {\bm k},\alpha | \frac{1}{E-{\cal H}_0 - {\cal V}\pm i\delta} 
| {\bm k}',\beta \rangle \big\rangle_{\rm imp} 
\end{eqnarray}
where $|{\bm k},\alpha\rangle \equiv |{\bm k}\rangle |\alpha\rangle$, $|{\bm k}\rangle$ 
a plane-wave state with momentum ${\bm k}$, and  $|\alpha\rangle$ either $s$ or $p$ orbital.  
${\cal H}_0$ is the non-interacting Hamiltonian in the clean limit and ${\cal V}$ 
denotes the impurity potential part in Eq.~(1) in the main text, i.e. 
${\cal V}=\sum_{\bm x} v_{s}({\bm x}) c^{\dagger}_{{\bm x},s}c_{{\bm x},s}
+ \sum_{\bm x} v_{p}({\bm x}) c^{\dagger}_{{\bm x},p}c_{{\bm x},p}$. 
The impurity averages are taken as follows,  
\begin{align}
&\big\langle v_{s}({\bm x}_j) v_{s}({\bm x}_{m}) \big\rangle_{\rm imp}  = 
\big\langle v_{p}({\bm x}_j) v_{p}({\bm x}_{m}) \big\rangle_{\rm imp} \nonumber  \\ 
& \ \  = \delta_{j,m} 
\Bigg(\int^{\frac{W}{2}}_{-\frac{W}{2}}   v^2 dv\Bigg) /
\Bigg(\int^{\frac{W}{2}}_{-\frac{W}{2}} dv\Bigg)  =  \delta_{j,m}  \frac{W^2}{12} \equiv 
\delta_{j,m} K. \nonumber
\end{align} 
with $\langle v_{p}({\bm x}_j) v_{s}({\bm x}_{m}) \rangle_{\rm imp}  = 0$. 
Within the self-consistent Born approximation, 
the Green function after quenched 
average takes a form of  $[{\bm G}_{\pm}({\bm k},{\bm k}')]_{\alpha,\beta} 
= \delta_{{\bm k},{\bm k}'} [{\bm G}_{\pm}({\bm k})]_{\alpha,\beta}$ where the 2 by 2 
matrix ${\bm G}_{\pm}({\bm k})$ is given by itself,  
\begin{eqnarray}
{\bm G}^{-1}_{\pm}({\bm k})  = {\bm G}^{-1}_{0,\pm} ({\bm k}) 
- \frac{K}{2} \frac{1}{N} \sum_{\bm q} \sum_{\mu=0,3} {\bm \sigma}_{\mu} 
{\bm G}_{\pm} ({\bm q}) {\bm \sigma}_{\mu}\, \nonumber  
\end{eqnarray}  
with $N$ the number of sites.
${\bm G}^{-1}_{0,\pm}({\bm k})$ denotes a non-interacting Green function, 
${\bm G}^{-1}_{0,\pm}({\bm k})\equiv (\omega\pm i \delta) - {\bm H}({\bm k})$ 
with Eq.~(\ref{eq1}). The approximation takes into account all the 
`non-crossing' diagrams, while `crossing' diagrams (Feynman diagrams 
containing crossings of two impurity potential lines) are smaller than 
non-crossing diagrams by an additional factor proportional to $\tau^{-1}$. 
Thus, the approximation is justified in the limit of long life-time quasiparticle 
($\tau^{-1} \rightarrow 0$), which is indeed the case with the zero-energy 
state ($E=0$) in the renormalized Weyl semimetal (WSM) phase (see below). 

The averaged Green function is expanded in terms of the Pauli matrices 
with complex-valued coefficients,   
\begin{eqnarray}
{\bm G}^{-1}_{+} ({\bm k}) = b_{0} \sigma_{0} 
- \sum_{j=1}^3 b_{j}({\bm k}) \sigma_{j}. \label{Greenfn}
\end{eqnarray}
The symmetric treatment of the impurity average ( 
$\langle v_{s} v_{s}\rangle_{\rm imp} = \langle v_{p} v_{p} \rangle_{\rm imp}$ and 
$\langle v_{s} v_{p} \rangle_{\rm imp} =0$) enables 
$b_{1}({\bm k})=a_{1}({\bm k})$ and $b_{2}({\bm k}) = a_{2}({\bm k})$. 
$b_{0}= E+i\delta - \gamma_0$ and $b_3({\bm k}) = a_3({\bm k}) + \gamma_3$ 
with ${\bm k}$-independent complex-valued coefficients $\gamma_{0}$ and $\gamma_{3}$. 
These two are determined by the following coupled self-consistent equations;
\begin{align}
\gamma_0 &= \frac{K}{N} \sum_{\bm k} \frac{\gamma_0 - E_{+} }{
a^2_1({\bm k}) + a^2_2({\bm k}) + (a_3({\bm k}) + \gamma_3)^2  
- (E_{+} -\gamma_0)^2}, \label{g0} \\
\gamma_{3}  & = - \frac{K}{N} \sum_{\bm k} \frac{a_3({\bm k}) + \gamma_3} 
{a^2_1({\bm k}) + a^2_2({\bm k}) + (a_3({\bm k}) + \gamma_3)^2  
- (E_{+} -\gamma_0)^2}. \label{g3} 
\end{align}
with $E_{+}= E+i\delta$

For the zero-energy state ($E=0$), Eqs.~(\ref{g0}) and (\ref{g3}) reduce to a 
simpler set of coupled equations. The simpler set of equations are only for ${\rm Im}\gamma_0=\tau^{-1}$ 
and ${\rm Re}\gamma_3 \equiv M_3$ while ${\rm Re} \gamma_0=0$ and 
${\rm Im}\gamma_3 = 0$; 
\begin{align}
\frac{1}{\tau} &= \frac{1}{\tau} \frac{K}{N} \sum_{\bm k} 
\frac{1}{\overline{E}^2({\bm k}) + \tau^{-2}}, \label{g0a} \\
M_3 &= - \frac{K}{N} \sum_{\bm k} 
\frac{a_3({\bm k}) + M_3}{\overline{E}^2({\bm k}) + \tau^{-2}},  \label{g3a} \\ 
\overline{E}^2({\bm k}) &= a^2_1({\bm k}) + a^2_2({\bm k}) + (a_3({\bm k}) + M_3)^2. 
\label{gsup}
\end{align}
For weaker disorder case, the first equation can be satisfied only by putting 
$\tau^{-1}=0$. The solution with $\tau^{-1}=0$ (zero-energy state with 
infinite life time) corresponds to either the renormalized Weyl 
semimetal (WSM) phase or gapped Chern insulator (CI) phase.
A critical disorder strength $K_{c}$ above which the coupled equations  
can have a solution with finite life time ($\tau^{-1} \ne 0$) 
is determined by the following gap equation~\cite{24,27},       
\begin{eqnarray}
1 =  \frac{K_c}{N}
\sum_{\bm k} \frac{1}{\overline{E}^2({\bm k})}  \label{g0aaa}
\end{eqnarray} 
where $M_3$ in $\overline{E}({\bm k})$ in the right hand side 
is obtained from Eq.~(\ref{g3a}) with 
$\tau^{-1}=0$. In fact, for $K>K_c$, finite $\tau$ 
can satisfy    
\begin{eqnarray}
1 =  \frac{K}{N}
\sum_{\bm k} \frac{1}{\overline{E}^2({\bm k}) + \tau^{-2}}.  \label{g0aa}
\end{eqnarray}
The finite-$\tau$ solution (zero-energy state 
with finite life time) corresponds to the diffusive metallic (DM)
phase~\cite{24,27}. 

For $K<K_c$, we solve Eq.~(\ref{g3a}) with $\tau^{-1}=0$ 
in favor for $M_3$ (renormalized WSM or CI phases). 
For $K\ge K_c$, we solve Eqs.~(\ref{g3a}) and (\ref{g0aa}) 
for $M_3$ and $\tau$ (DM phase).  In either cases, 
an electronic dispersion is renormalized by the disorder-induced mass term 
$M_3$. In the present set of parameters with $\alpha=-\frac{1}{2}$, $M_3$ 
becomes positive. Fig.~\ref{pd} suggests that the positive mass term with $\alpha=-\frac{1}{2}$ 
drives the CI phase and the region-III WSM phase (with three pairs of Weyl points) 
near $|\beta|=0.5$ into the region-II WSM phase (with two pairs of Weyl points). 
Namely, for the CI phase ($\beta>-\frac{1}{2}=\alpha$), larger positive mass $M_3$ 
realizes $\alpha+M_3>\beta$, where two pairs of Weyl points 
are created at ${\bm k}=(\pi,0,\pi\pm \overline{k}_0)$ and $(0,\pi,\pi\pm \overline{k}_0)$ with 
\begin{eqnarray}
\overline{k}_0 \equiv \cos^{-1}[\frac{\alpha+M_3}{\beta}].  \label{k0}
\end{eqnarray}
For the region-III WSM phase ($\beta<-\frac{1}{2}=-1-\alpha$), a positive mass term can satisfy 
$1+\alpha+M_3\ge -\beta$, so that a pair of two Weyl points at 
${\bm k}=(\pi,\pi,\pm \overline{k}_1)$ annihilate with each other with 
\begin{eqnarray}
\overline{k}_1 \equiv \cos^{-1}[-\frac{1+\alpha+M_3}{\beta}]. \label{k1}
\end{eqnarray} 
By solving Eq.~(\ref{g0aaa}), we determine a phase boundary 
between diffusive metallic phase and renormalized 
WSM phases. By solving Eq.~(\ref{g3a}) with $\tau^{-1}=0$ in favor for $M_3$, 
we determine a 
boundary between the region-II WSM phase and Chern insulator from $\alpha + M_3=\beta$, 
and a boundary  
between the region-II WSM and region-III WSM phases from $1+\alpha+M_3=-\beta$. This 
completes a phase diagram obtained from the self-consistent Born analysis  
(inset of Fig.~1 in the main text). 
By solving Eqs.~(\ref{g0a}) and (\ref{g3a}) for $M_3$ and $\tau$,   
we also obtain Eqs.~(\ref{k0}) and (\ref{k1}); ``renormalized'' locations of the Weyl points 
in the momentum space. 
By substituting $\overline{k}_0$ and $\overline{k}_1$ into Eq.~(3)  
in the main text, we calculate the surface conductance due to the ``renormalized'' chiral 
Fermi arc, $\Delta G \equiv G_o - G_p$, as a function of the disorder strength 
(Fig.~5 in the main text). For $K\le K_c$, $\Delta G$ thus obtained 
quantitatively reproduces the numerics. 

\begin{figure}
\centering
\includegraphics[width=0.85\linewidth]{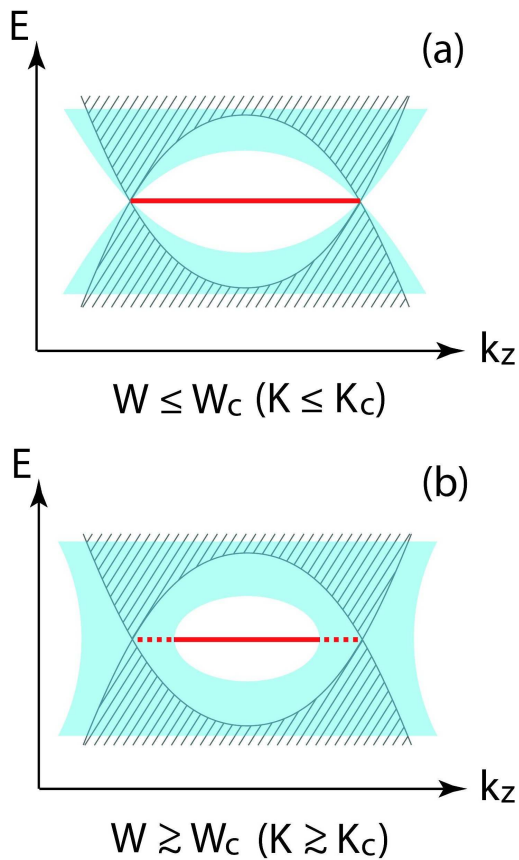}
\caption{(color online) Schematic pictures of an energy-momentum (surface crystal 
momenta) dispersion for bulk states with a pair of two Weyl nodes (black) and for corresponding 
chiral surface Fermi arc state at $E=0$ (a red line). (a) $K\le K_c$ and (b)  
$K\gtrsim K_c$. As in Fig.~\ref{CFA}, the surface crystal momenta comprise of two momenta, 
$k_x$ and $k_z$, while $k_x$ is omitted for simplicity; the surface Fermi arc state 
has a chiral dispersion along the $k_x$ direction. Note also that we only show one pair 
of two Weyl nodes for clarity. The disorder endows bulk states with 
a finite life time (line broadening), whose effect is represented by a green hatch in the figure. 
Within the green hatch, the crystal momentum is not well-defined even after quenched 
averaging of disorder potentials; any two states within the hatch are mixed with each other 
and indistinguishable due to the disorder. (a) for $K\le K_c$, an inverse of the life time 
(line broadening) for the two Weyl nodes is zero, so that {\it all} the chiral arc states at $E=0$ 
connecting these two nodes provide protected surface conductions. 
(b) for $K\gtrsim K_c$, those chiral fermi arc states at $E=0$ which are close to 
the two ends of the arc (those red dotted lines covered by the green hatch) are mixed with 
bulk states due to the finite line broadening of the two Weyl nodes; they will not contribute to 
surface conductions. Meanwhile, those arc states at $E=0$ which are well 
separated from the two ends (red solid line not covered by the green hatch) are 
{\it not} mixed with bulk states, providing protected surface conductions. }
\label{liushang}
\end{figure}

Importantly, the SCB analysis shows that paired two 
Weyl nodes do {\it not} annihilate with each other at the 
semimetal-metal quantum phase transition point ($K=K_c$). Thereby, 
the associated chiral surface Fermi arc state also {\it does}  
survive at the transition point. Instead of annihilating with each other, 
the two nodes acquire a finite line broadening {\it individually}; 
an inverse of their life time ($\tau^{-1}$) evolves continuously from 
zero to a finite value for $K\ge K_c$ (as described above). Due to this line 
broadening of the Weyl nodes, those chiral surface states 
{\it near} the two ends of the arc start being mixed with bulk states, losing 
their protected surface conduction nature. On the one hand, those chiral  
surface states {\it far away from} the two ends still remain intact for $K \gtrsim K_c$, 
because they are well separated from bulk states (Fig.~\ref{liushang}). 
The line broadening of the Weyl nodes evolve continuously, so that the latter surface states can still contribute to 
a residual surface conduction $\Delta G$ for those $K$ {\it moderately} greater than $K_c$. 
In fact, the numerics (Fig.~5 in the main text) shows that a surface conductance remains 
finite for $W\lesssim 2W_c$, while the DM phase itself ranges up to a much 
larger disorder strength. 

Note also that the persistence of paired two Weyl nodes at $K=K_c$ 
is consistent with the vanishing effective velocity $\overline v$ 
at the transition point (observed by the density of states analysis). 
The SCB analysis explains $\overline{v}=0$ at $K=K_c$ by an anomalous 
scaling of a renormalized chemical potential. Specifically, the SCB analysis shows 
that, for smaller $E$, the real part of $b_0$ in Eq.~(\ref{Greenfn}) is scaled with $E$ 
for $K<K_c$ and with $\sqrt{E}$ for $K=K_c$.

Within the self-consistent Born approximation, the density of states is proportional 
to an inverse of the life time $\tau$, 
\begin{align}
&\rho(E) \equiv - \frac{1}{\pi}\frac{1}{N} \bigg \langle 
{\rm Im}\!\ {\rm Tr} \Big[\frac{1}{E-{\cal H}_0 - {\cal V}+i\delta}\Big] \bigg \rangle_{\rm imp} \nonumber \\ 
& = - \frac{1}{\pi}\frac{1}{N} \sum_{\bm k} 
{\rm Im}\!\ {\rm tr}G_{+}(E,{\bm k}) = \frac{2}{\pi \tau K}. \label{dos} 
\end{align}
Numerical solutions of Eqs.~(\ref{g0}) and (\ref{g3}) dictate that 
$\tau^{-1}$ for smaller $E$ is proportional to $E^2$ in the renormalized WSM 
phases ($K<K_c$)~\cite{24,27},  
while being scaled to $\sqrt{E}$ at the quantum critical point ($K=K_c$). 
The solutions also find that the inverse of the life time of the zero-energy state 
in the DM phase ($K > K_c$) grows linearly in  
$K-K_c$~\cite{24,27}.  
$\tau^{-1} \propto E^2$ for $K<K_c$ is consistent with the 
behaviour of $\rho(E)$ for $W<W_c$ obtained from the kernel polynomial expansion 
(Fig.~3 in main text). $\tau^{-1} \propto \sqrt{E}$ at $K=K_c$ results in  
a dynamical exponent $z_{\rm scb}=2$, which is different from the exponent evaluated from the numerics 
($z\simeq 1.5$). $\tau^{-1} \propto K-K_c$ at $E=0$ 
for $K>K_c$ in combination with the dynamical exponent $z_{\rm scb}=2$ 
gives the other exponent $\nu_{\rm scb}=1$. Discrepancies of the critical exponents obtained 
from these two calculations stem from the neglected Feynman diagrams 
(`crossing' diagrams) in the self-consistent Born approximation.

The Einstein's relation relates the longitudinal electric conductivity $\sigma(E)$ with 
the diffusion constant $D(E)$ and the density of states $\rho(E)$, 
\begin{eqnarray}
\sigma(E) = e^2 D(E)\rho(E). \label{einstein}
\end{eqnarray}  
The diffusion constant $D(E)$ is given by mean-free length 
$l = \overline{v}\tau$ and life time (mean free time) $\tau$ as 
$D(E)=l^2/3\tau=\overline{v}^2\tau/3$, where $\overline{v}$ stands for an effective  
velocity of electrons. As shown above, the density of states $\rho(E)$ is inversely 
proportional to the life time within the self-consistent Born approximation. Thus, when 
the energy is set to the zero in the renormalized WSM phase, $\rho(E)$ vanishes as 
$1/\tau$, $D(E)$ diverges as $\tau$, and the conductivity remains constant; 
$\sigma(E=0) = 2\overline{v}^2/(3\pi K)$. This dictates that the conductivity 
at $E=0$ remains finite not only in the diffusive metallic phase but also in the renormalized 
WSM phase, while vanishing at the quantum critical point intervening these 
two phases where $\overline{v} \rightarrow 0$ (see the main text).

\subsection*{Chiral surface states in layered Chern insulator 
and in renormalized Weyl semimetal phase} 
\begin{figure}
\centering
\includegraphics[width=0.9\linewidth]{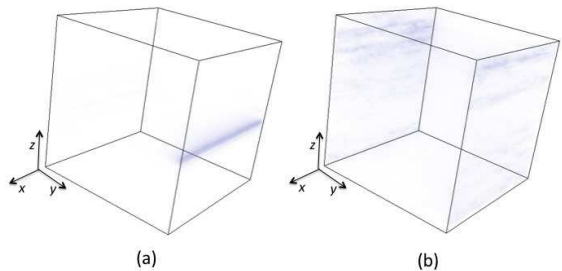}
\caption{(color online) Wave function density examples 
for layered Chern insulator (a) ($\beta=0.45, W=0.8, E_\nu=-0.00000859\cdots$), and
a wave function density for Weyl semimetal  (b) ($\beta=0.45, W=1.7, E_\nu=-0.0000688\cdots$).
In the former case,  the charge density appears one dimensionally along $x$ direction on the $x-z$ plane,
while in the latter case, the charge density appears two dimensionally on the $x-z$ plane.}
\label{tomi1}
\end{figure}

Eigenstates of Eq.~(1) in the main text for a given disorder realization are numerically 
calculated with the periodic boundary condition along the chiral  ($x$) 
and the stacking  ($z$) directions,
and open boundary condition for the other directions ($y$). 
System size $L=80$, and numerical calculation has been done by sparse matrix diagonalization algorithm of Intel MKL/FEAST.
Figure~\ref{tomi1} shows spatial density distributions of an eigenstate $\nu$,
\[
\rho_\nu(x,y,z)=|\psi_{\nu, s}(x,y,z)|^2+|\psi_{\nu,p}(x,y,z)|^2\,.
\]
We took the states $\nu$ which are closest to the  
zero energy ($E=0$) in the layered Chern insulator phase 
region (a)  ($|\beta|=0.45$ and $W=0.8$), and in the renormalized Weyl semimetal phase region (b)
($|\beta|=0.45$ and $W=1.7$). 
In both cases, the eigenstates are nearly localized at the boundaries 
($y=\pm L/2$), suggesting that zero-energy state or eigenstates 
close to the zero energy are usually chiral surface states. 
Figure~\ref{tomi1} also indicates that the chiral surface state in the renormalized WSM phase 
is spatially extended within surfaces, while the chiral state in the  
layered CI phase is extended only along the chiral direction and 
localized along the stacking direction.

\subsection*{Finite size scaling of the density of states }
\begin{figure}
\centering
\includegraphics[width=0.9\linewidth]{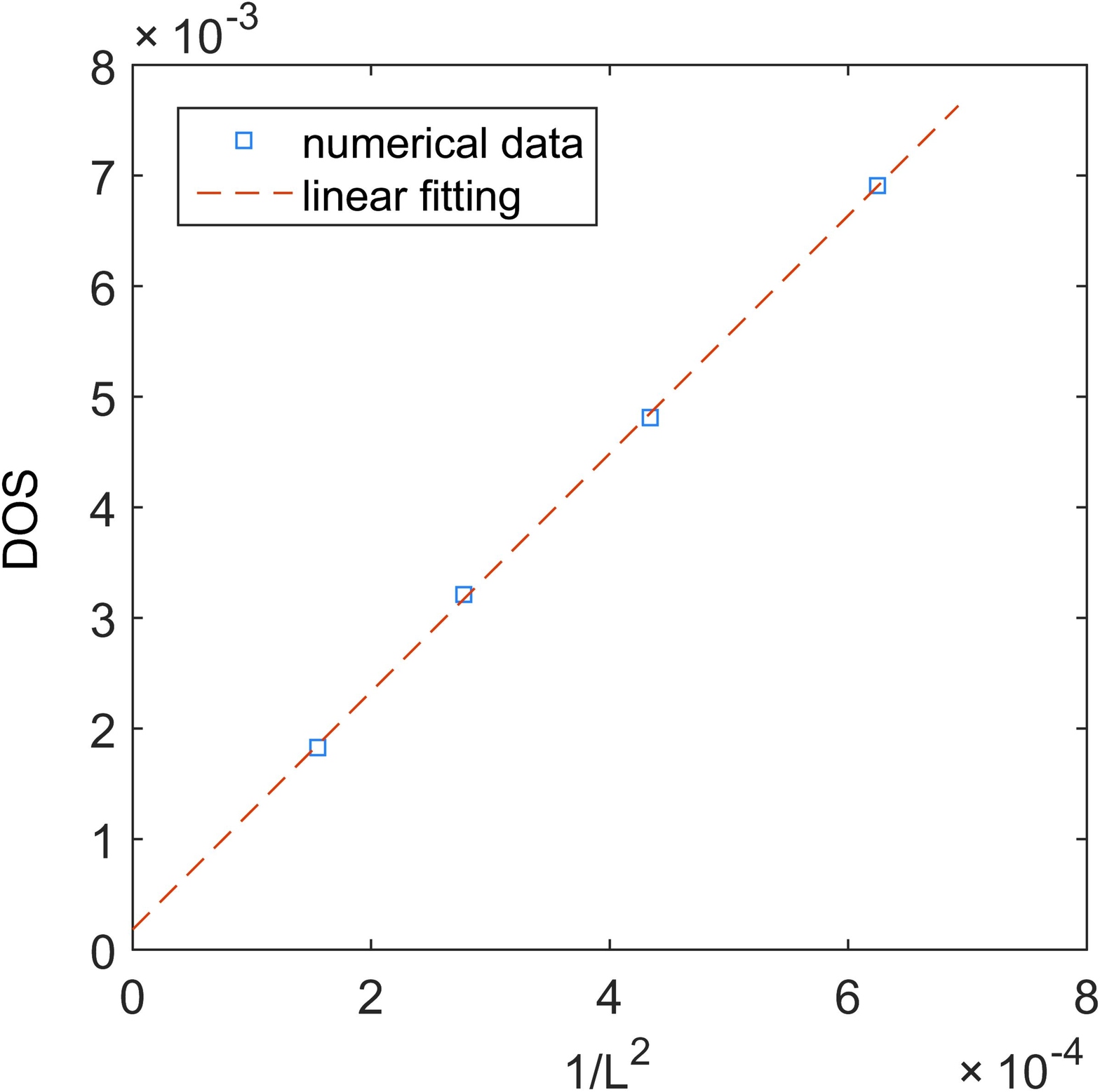}
\caption{(color online) Finite size scaling of the DOS at $E=0$ and $W=1.5$. The blue squares denote
numerical data of the DOS for four different system sizes, and the red dashed line is the linear fit to $1/L^2$.}
\label{fitting}
\end{figure}

\begin{figure}
\centering
\includegraphics[width=0.9\linewidth]{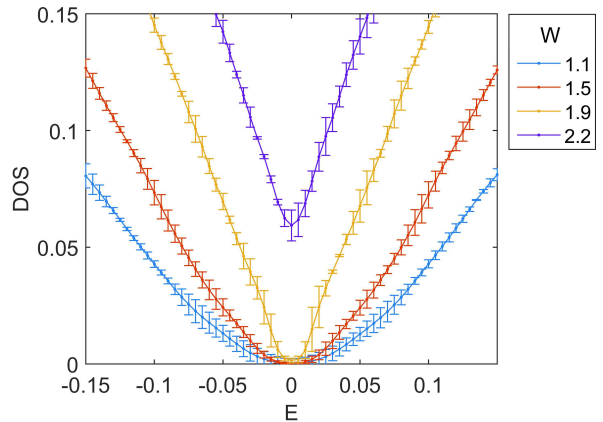}
\caption{(color online) The DOS from the finite size scaling for typical values of
$E$ and $W$. The 95\% confidence intervals are indicated as error bars .}
\label{errorDOS}
\end{figure}

The density of states (DOS) is calculated for different system sizes ($L=40,~48,~60,~80$) in terms of
the kernel polynomial expansion method \cite{18}.
The calculated DOS is fitted with the finite size scaling form,
$\rho_{L}(E,W)=a(E,W)+b(E,W)/L^2$ \cite{25}, which gives the DOS in the thermodynamic 
limit as the intercept $a(E,W)$.  An example of the fitting 
is shown in Fig.~\ref{fitting} (for $\beta=-0.6$, $E=0.0$ and
$W=1.5<W_c$). The error bar obtained from this finite-size-scaling fit is shown in
Fig.~\ref{errorDOS} for typical values of $E$ and $W$ with $\beta=-0.6$.
The error is estimated with 95\% confidence (2$\sigma$).
Since the fitting has only two degrees of freedom, the estimated error bar is
relatively large.

\subsection*{Parabolic fitting and the velocity}
\begin{figure}
\centering
\includegraphics[width=0.9\linewidth]{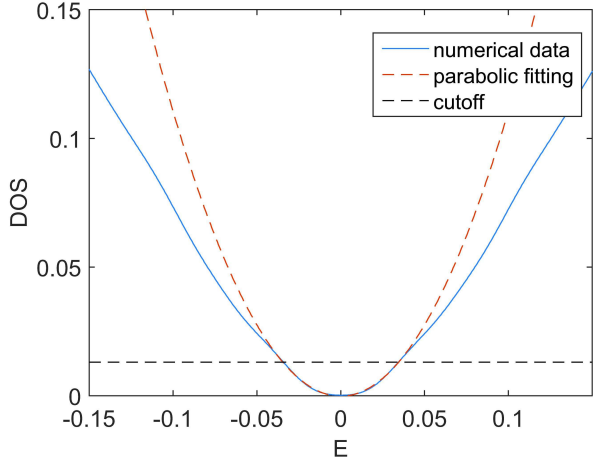}
\caption{(color online) A parabolic fitting of the DOS at $W=1.5$ and $\beta=-0.6$.
The blue solid line is from numerical data and the red dashed line is a $E^2$-fitting curve. 
To excludes
those data points away from $E=0$, we introduce a cutoff for the DOS ($\rho<0.013$),
which is depicted by a black dashed horizontal line. }
\label{parabolic}
\end{figure}

In the renormalized WSM phase ($W<W_c$), the averaged velocity of the Weyl point is obtained
from a fitting of the DOS near $E=0$ by a parabolic function $\rho(E)=a+bE^2$.
Fig.~\ref{parabolic} shows an example of the fitting ($\beta=-0.6$, $W=1.5<W_c$).
On increasing $|E|$, the DOS curve changes from the $E$-square behavior to
a $E$-linear behaviour. When $W$ becomes closer to $W_c$ from below,
the $E$-square region becomes narrower in energy, indicating the reduction of the
velocity. To estimate the curvature only around $E=0$ from the fitting, we first observe
this crossover by eye and then introduce a cutoff for $\rho(E)$ to exclude those data
points in the $E$-linear region (For example, $\rho(E)<0.013$ in Fig.~\ref{parabolic}).

\subsection*{Determination of the critical point}
\begin{figure}
\centering
\includegraphics[width=0.9\linewidth]{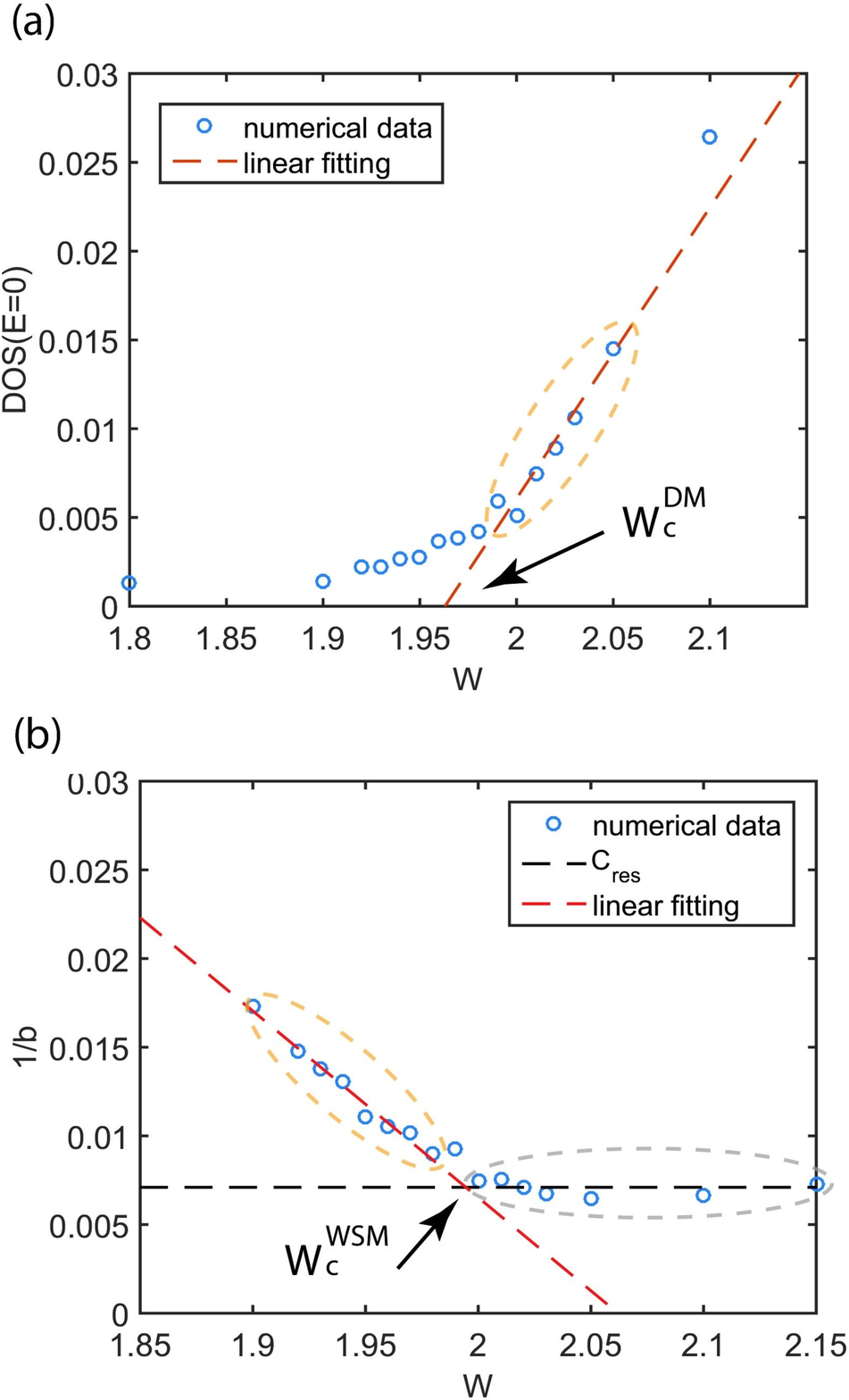}
\caption{(color online) (a) Determination of the critical disorder $W_c^{DM}$ from the density of states (DOS) at $E=0$.
We selected a group of data points in the diffusive metal (DM) side and tried a linear fitting (red dashed line).
The intercept with the $\rho(0)=0$ axis is identified with the critical disorder strength $W^{DM}_c$.
(b) Determination of the critical disorder $W^{WSM}_c$ from the averaged velocity of the Weyl cone $1/b$.
The data points in the DM side, which are near the WSM phase, take approximately small but constant value $C_{\text{res}}$
(black dashed line). We applied a linear fitting to a group of data points in the WSM side (red dashed line).
The intercept with $1/b=C_{\text{res}}$ is regarded as the critical disorder strength $W^{WSM}_{c}$, below which
the averaged velocity starts to take a finite value.}
\label{critical_point}
\end{figure}

The phase boundary between WSM and DM is determined by an onset of finite DOS at $E=0$
and also by an onset of finite averaged velocity of the Weyl cone.
The DOS at $E=0$ vanishes for smaller disorder strengths. It starts to
take a finite value when $W$ is greater than a critical disorder strength.
To determine this critical disorder strength without ambiguity, we first select a set of data points
in the $\rho(0)-W$ figure (typically five to eight data points), which are in the DM region within the
finite-size-scaling error bar (explained above), but which are sufficiently close to the WSM region.
We applied a $W$-linear fitting for these data points, to regard the intercept of the fitting curve
with the $\rho(0)=0$ axis as the critical disorder strength 
$W_c^{DM}$ (Fig.~\ref{critical_point}(a)). 

On increasing the disorder strength
in WSM phase, the inverse of the parabolic coefficient $1/b$
decreases continuously, only to converge to a small but non-zero constant value $C_{\text{res}}$
in DM phase side. The small non-vanishing $C_{\text{res}}$ is due to non-zero error bar
stemming from the finite size scaling and the finite truncation of KPE. 
To derive a critical disorder strength as an onset disorder strength of
finite averaged velocity, we regard this residual value $C_{\text{res}}$ to be effectively zero and determine
the critical strength from an intersect of a curve for $1/b$ as a function of $W$ with $1/b=C_{\text{res}}$.
To be more specific, we firstly fit out $C_{\text{res}}$ from a set of those data
points in the DM region which are close to the WSM region (data points encompassed by a grey
dotted circle in Fig.~\ref{critical_point}(b)). We then collects another group of data points from
the WSM region which are sufficiently close to the DM region (data points encompassed by an orange-colored
dotted circle in Fig.~\ref{critical_point}(b)). As above, we applied a $W$-linear fitting for the
latter group of the data points, and regard its intercept with $1/b=C_{\text{res}}$ as the critical
disorder strength $W_c^{WSM}$ (Fig.~\ref{critical_point}(b)).

\subsection*{Single parameter scaling}
\begin{figure}[h]
\centering
\includegraphics[width=0.9\linewidth]{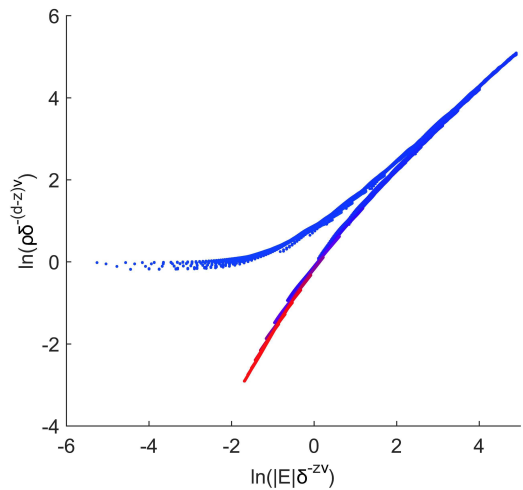}
\caption{(color online) A single parameter scaling of the density of states for $|\beta|=0.6$ and
$W$ near the critical point with $\rho(E)>0.02$. All the data points collapse onto two
distinct branches which correspond to a scaling  function for  WSM phase
(lower branch) and that for DM phase (upper branch), respectively.}
\label{scaling}
\end{figure}

According to a scaling argument\cite{25,29}, the density
of states near the zero energy state follows scaling functions near the quantum critical point;
\begin{equation}
\rho(E) = \xi^{z-d} F(|E| \xi^z) = \delta^{(d-z)\nu}f(|E|\delta^{-z\nu}),
\label{scalinglaw}
\end{equation}
where a characteristic length $\xi$ scaled with the distance measured from
the critical disorder strength, $\delta\equiv |W-W_c|/W_c$ with
the critical exponent $\nu$ as $\xi\propto \delta^{-\nu}$. $z$ denotes the dynamical exponent and
the spatial dimension $d$ is 3 in the present case. The critical disorder strength $W_c$ is chosen to
be the mean value between $W^{WSM}_c$ and $W^{DM}_c$, since these two coincides with each
other within the error bar. A scaling function $f(\cdots)$ for the WSM phase
and that for the DM phase are generally different from each other, while the critical exponents
($\nu$ and $z$) are universal. At the critical point, the $\delta$-dependences in
Eq.~(\ref{scalinglaw}) should cancel exactly, leading to
$\rho(E)\propto\delta^{(d-z)\nu}(|E|\delta^{-z\nu})^{(d-z)/z}=|E|^{(d-z)/z}$. The calculated
DOS at $W=W_c$ is nearly linear in $|E|$, suggesting that 
$z \simeq d/2$~\cite{30,34}. A comparison
between the numerical data and the scaling form gives $z=1.53\pm 0.03$. In WSM phase,
the DOS near $E=0$ is scaled with $|E|^{d-1}$, so that the scaling form requires that
its coefficient should be scaled with $\rho(E) \propto |E|^{d-1}\delta^{-d(z-1)\nu}$. Fitting of 
an coefficient of $E^2$ from Fig.~3b in the main text along this scaling form, 
i.e. $1/b \propto \delta^{-d(z-1)\nu}$,  gives out $d(z-1)\nu=1.24\pm 0.06$. In DM phase, the DOS at $E=0$
is scaled with $\rho(0) = \delta^{(d-z)\nu}f(0)$, along which the DOS at $E=0$ from Fig.~3b in the main text is fitted, giving $(d-z) \nu=1.47\pm 0.06$. 
These two fittings in combination with $z=1.53\pm 0.03$
give us
\begin{eqnarray}
\nu^{WSM}&=&0.78\pm 0.08,\\
\nu^{DM}&=&1.00\pm0.13,
\end{eqnarray}
respectively. Though they do not coincide exactly, the discrepancy is at
the same order of our estimated error bar, giving a weighted average $\nu$ as
\begin{equation}
\nu=0.84\pm0.10.
\end{equation}
In fact, with this critical exponent and $z=1.53\pm 0.03$ in hand,
we can successfully fit all the data points of our DOS for $\rho(E)>\rho_c$ into two
scaling functions, provided that $\rho_c$ is chosen on the same order of
error bars stemming from the finite size scaling (see above). 
As shown in a log-log plot of $\rho\delta^{-(d-z)\nu}$ versus 
$|E|\delta^{z\nu}$ (Fig.~\ref{scaling}), all the data points collapse 
into two branches, where one branch (lower branch) corresponds to a scaling 
function for the WSM phase and the other (upper branch) to a scaling function
for the DM phase respectively.  With these results, we conclude that the critical nature of 
the WSM-DM phase transition in the unitary class follows the
scaling law of Eq.\eqref{scalinglaw} with $z=1.53\pm 0.03$ 
and $\nu=0.84\pm 0.1$.

\begin{figure}[h]
\centering
\includegraphics[width=0.9\linewidth]{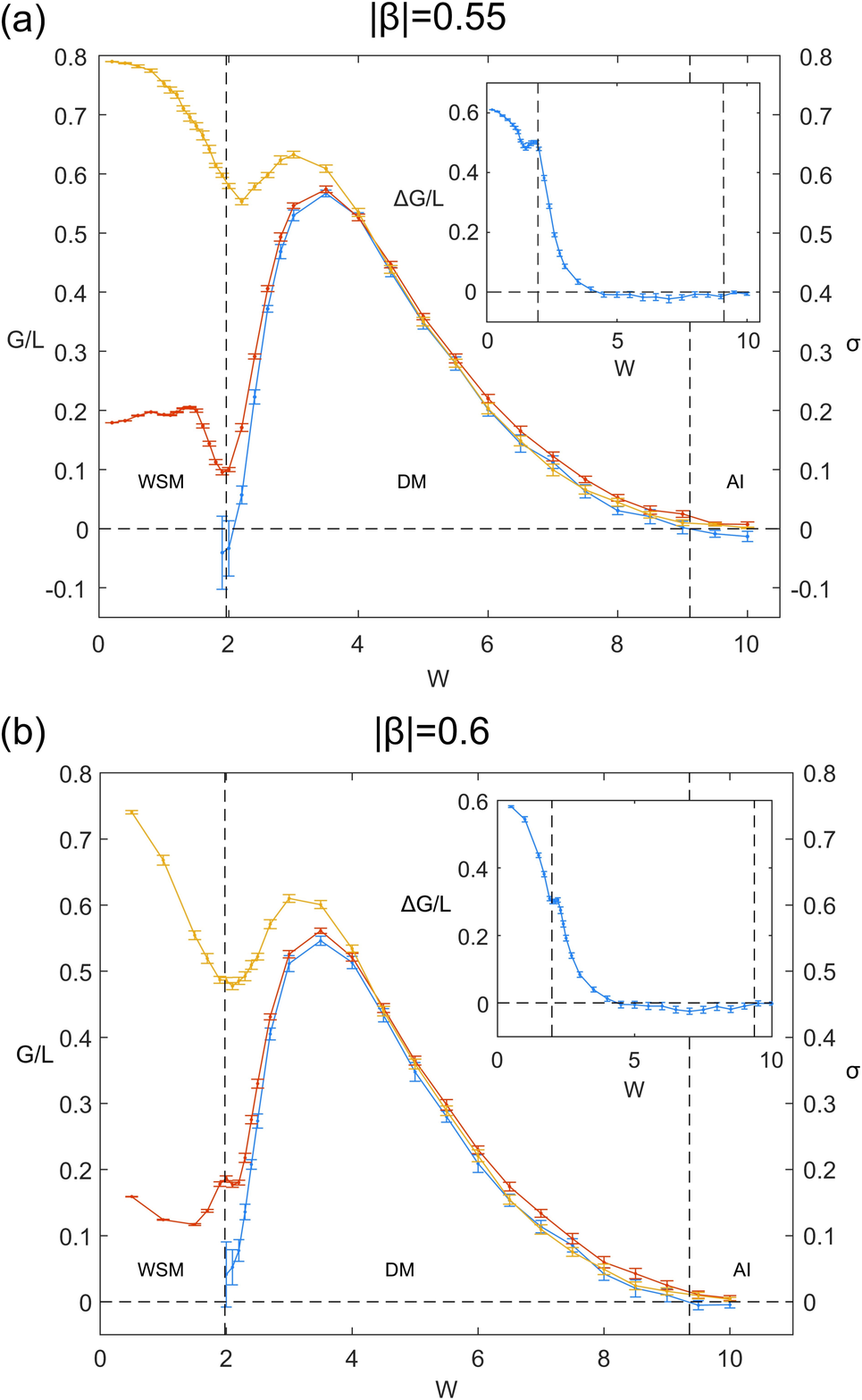}
\caption{(color online) In-plane ($x$-dir) bulk conductivity and conductance of the cubic system
($L^3$) at $|\beta|=0.55$ (left) and $0.6$ (right) as a function of the disorder strength $W$.
Along the other two spatial directions ($y$ and $z$-dir), we impose either periodic boundary 
condition $G^p$ (red points with lines) or open boundary condition $G^o$ (yellow points with lines) 
with $L=30$. The bulk conductivity $\sigma_b$ is obtained from the finite size scaling of the conductance
with periodic boundary condition (blue points with line). The two vertical dotted lines
(from left to right) stand for WSM-DM and DM-AI transition points respectively. 
The DM-AI transition point is determined by the localization length analysis, 
while the WSM-DM transition point is determined by the DOS analysis. (Inset) $\Delta G \equiv G^o-G^p$ 
for $|\beta|=0.55$ (left) and $0.6$ (right). }
\label{cond_06}
\end{figure}

\subsection*{Conductance and conductivity at $|\beta|=0.55,0.6$}
The two-terminal conductance and conductivity for $|\beta|=0.55$ and $0.6$ are shown in Fig.~\ref{cond_06}.
The conductance along the $x$ direction (in-plane direction) are calculated for various system size
($L=6,8,10,\cdots,30$), with periodic ($G^p$) or open boundary condition ($G^o$) in the $y$ and $z$ directions.
We prepared 20 uncorrelated disorder realizations, over which the calculated
conductance are averaged. The in-plane bulk conductivity $\sigma_b$ is obtained from $G^p$ by the linear
fitting ($G^p=\sigma_bL+c$).  We found that a finite bulk conductivity in the DM phase continuously
reduces to zero toward the WSM-DM transition point. This behaviour is consistent with
the self-consistent Born analyses \cite{27} and a generalized Wegner's
scaling relation \cite{32,25}. 

Note also that the two-terminal conductance in the WSM region for $|\beta|=0.55$ and $0.6$
strongly oscillate as a function of the cubic system size $L$, while the conductance
in the WSM phase for $|\beta|=0.45$ and $0.50$ show much less prominent oscillation.
The larger oscillations for $|\beta|=0.55$ and $0.6$ make it hard to determine the bulk 
conductivity in the WSM phase with smaller error bar (not shown in Fig.~\ref{cond_06}).
The oscillation is due to (i) the very long mean-free path in
the WSM phase and/or due to (ii) a mismatch between $\frac{2\pi N}{L}$ ($N$ arbitrary
integer less than $L$) and $\overline{k}_{0}$ (or $\overline{k}_1$) (the renormalized positions
of the Weyl cones along the $k_z$ axis).


\end{document}